\documentclass[]{spie}  

\usepackage{amsmath,amsfonts,amssymb}
\usepackage{graphicx}
\usepackage[colorlinks=true, allcolors=blue]{hyperref}

\title{Asgard/NOTT: water vapor and CO$_2$ atmospheric dispersion compensation system}

\author[a]{Romain Laugier}
\author[a]{Denis Defrère}
\author[b]{Michael Ireland}
\author[a]{Germain Garreau}
\author[c]{Olivier Absil}
\author[d]{Alexis Matter}
\author[d]{Romain Petrov}
\author[d]{Philippe Berio}
\author[e]{Peter Tuthill}
\author[a]{Marc-Antoine Martinod}
\author[f]{Lucas Labadie}
\affil[a]{Institute of Astronomy, KU Leuven, Celestijnenlaan 200D, 3001 Leuven, Belgium}
\affil[b]{The Australian National University, Australia}
\affil[c]{Liège Univ., Belgium}
\affil[d]{Université Côte d'Azur, Observatoire de la Côte d'Azur, CNRS, Laboratoire Lagrange, France}
\affil[e]{The University of Sydney, Australia}
\affil[f]{Univ. zu Köln, Germany}

\authorinfo{Further author information: (Send correspondence to R.L.)\\R.L.: E-mail: romain.laugier@kuleuven.be\\}

\begin{document} 
\maketitle

\begin{abstract}
The direct detection of exoplanets and circumstellar disks is currently limited by a combination of high contrast and small angular separation. At the scale of single telescopes, these limitations are fought with coronagraphs, which remove the diffracted light from the central source. To obtain similar benefits with interferometry, one must employ specialized beam-combiners called interferometric nullers.
Nullers discard the on-axis light and part of the astrophysical information to optimize the recording of light present in the dark fringe of the central source, which may contain light from circumstellar sources of interest. Asgard/NOTT will deploy an advanced beam-combination scheme offering favorable instrumental noise characteristics when the inputs are phased appropriately, although this tuning will require a specific strategy to overcome the resulting degeneracy.
Furthermore, this must bring the phase of the incoming light to a good accuracy across the usable spectrum. Since the fringe-tracker operates at different wavelengths, it can only sense part of the offending errors, and we discuss the measurement of these errors with the science detector.
NOTT operates in the L band and suffers from various effects such as water vapor, which has already been experienced with N-band nullers (Keck Interferometer Nuller, Large Binocular Telescope Interferometer). This effect can be corrected with prisms forming a variable thickness of glass and an adjustment of air optical path.
Moreover, observations in the L band suffer from an additional and important chromatic effect due to longitudinal atmospheric dispersion coming from a resonance of carbon dioxide at 4.3µm that is impractical to correct with glass plates because of its non-linear wavelength dependency. To compensate for this effect efficiently, a novel type of compensation device will be deployed leveraging a gas cell of variable length at ambient pressure.
After reviewing the impact of water vapor and CO$_2$, we present the design of this atmospheric dispersion compensation device for Asgard/NOTT and describe a strategy to maintain this tuning on-sky.

\end{abstract}


\section{Introduction}\label{sec:intro}
    The direct detection of exoplanets and exozodiacal disks in the infrared requires high angular resolution and precise wavefront. When the wavefront and is not sensed at the scientific wavelengths, chromatic errors occur in the presence of atmospheric dispersive elements such as water vapor and $CO_2$. This limitation is well know in phase-referenced interferometers such as the Keck Interferometer Nuller (\cite{Colavita2004}\cite{Koresko2006}), Large Binocular telescope interferometer (\cite{Defrere2014}), and more recently GRA4MAT (Woillez et al. in press). This is also a well know problem for ELT instruments such as METIS (\cite{Kendrew2008}, \cite{Absil2022}).
    
\begin{figure}
        \centering
        \includegraphics[width=0.7\textwidth]{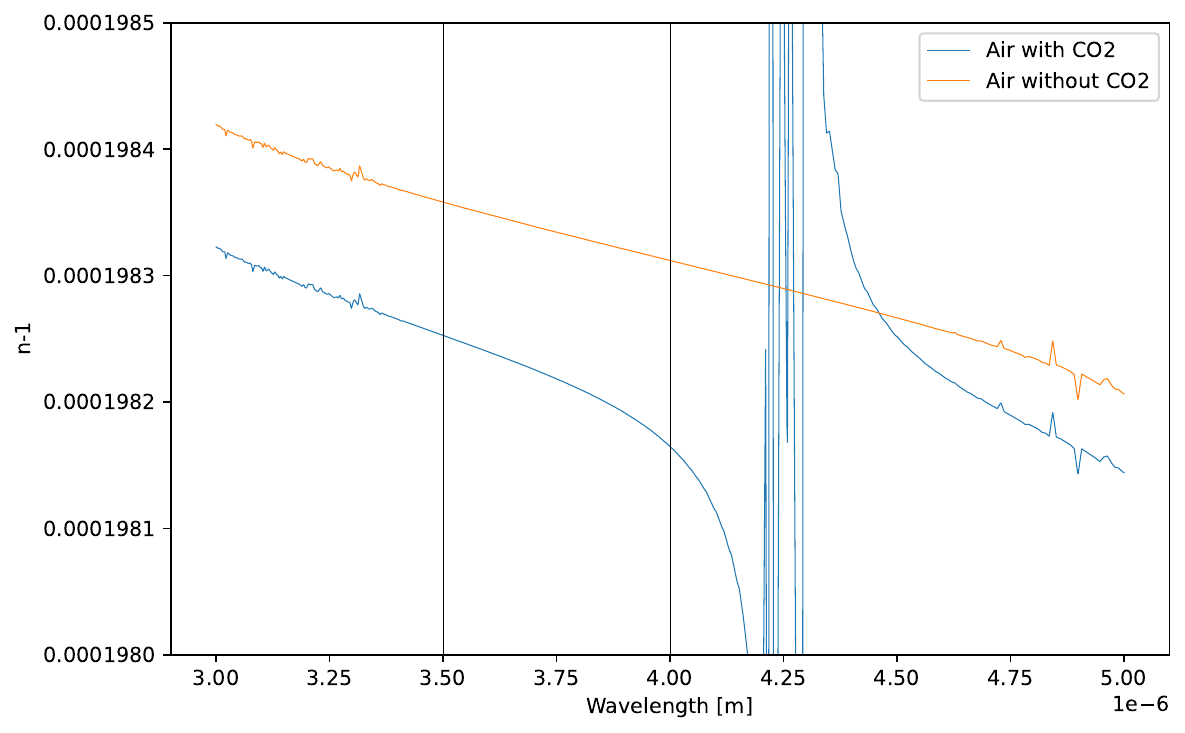}
        \caption{Refractive index of air computed based on models outlined in \cite{Mathar2007}, with and without the contribution of CO$_2$. The vertical blakc lines indicate the wavelength range of NOTT.}
        \label{fig:air_index}
    \end{figure}

    In the cas of nullers, the combination of input beams at an accurate phase setting to mitigate the on-axis starlight and obtain intended instrument behavior. In this work, we focus on the case of Asgard/NOTT\cite{Defrere2018}\cite{Defrere2022a}, the L-band nuller for Asgard, an instrument suite proposed for the visitor focus of the VLTI\cite{Martinod2023}. The error budget of NOTT \cite{Garreau2022, Sanny2022a, Garreau2024} allows for up to $10^{-2}$ raw contrast from the static instrument errors, which requires to correct phase better than $0.1$ rad across the L band. While the effect of water vapor was expected in the initial design of NOTT based on the experience from N-band nulling interferometers, the negative impact of CO$_2$ (see Fig. \ref{fig:air_index}) was not anticipated and requires a dedicated correction system. We present such a system in this paper.

\section{Optimal corrections}\label{sec:optim}
    The usual approach for longitudinal dispersion compensation is the one proposed by Tango \cite{Tango1990}, which uses a minimization of the spectral slope of the phase. For nulling interferometers, this is critical since nulling the on-axis light requires precise phase control. We use the approach of Koresko \cite{Koresko2006} to project the chromatic perturbation of the atmosphere onto the parameter space of the correcting materials. Using this approach, glasses like ZnSe and CaF$_2$ can offer near-fixed slope of phase which can be adjusted with moving prisms producing a variable thickness, such as was used for the Keck nuller\cite{Colavita2004}.
    
\begin{figure}
        \centering
        \includegraphics[width=0.9\textwidth]{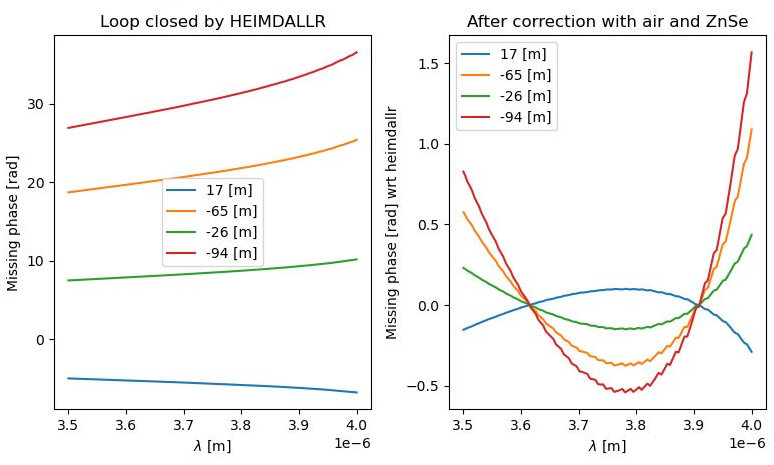}
        \caption{Phase residual after the fringe tracking by HEIMDALLR in the K band before (left) and after (right) correction by the NOTT delay lines and an adjusted thickness of glass.}
        \label{fig:path_imbalance}
    \end{figure}

    We used SCIFYsim \cite{Laugier2023} to simulate the performance of the system in the worst operating conditions, which are obtained observing a target at $\approx 30 \deg$ elevation with 30\% relative humidity, leading  for some baselines and pointings to a compensation by the delay line of $\approx 100 m$. Figure \ref{fig:path_imbalance} shows the resulting phase offset given by closing the loop with the Asgard fringe tracker (i.e. HEIMDALLR\cite{Taras2024a}) fringe tracker in the K band before compensating with the air delay lines of NOTT, and after compensation by air and glass (here ZnSe). The effect of most components of the atmosphere can be corrected with such variable thickness glass plate. However, CO$_2$ has an asymmetric stretch vibration absorption line at 4.3 µm that creates a strong curvature to the optical index of the ambient air. For our worst case scenario, a 100 m relative air imbalance leads to 0.5 rad phase error, even after correction by glass. Combinations of multiple glasses were inefficient at correcting such curved error.

\section{Gas system design}\label{sec:gazdesign}    

    \begin{figure}
        \centering
        \includegraphics[width=0.9\textwidth]{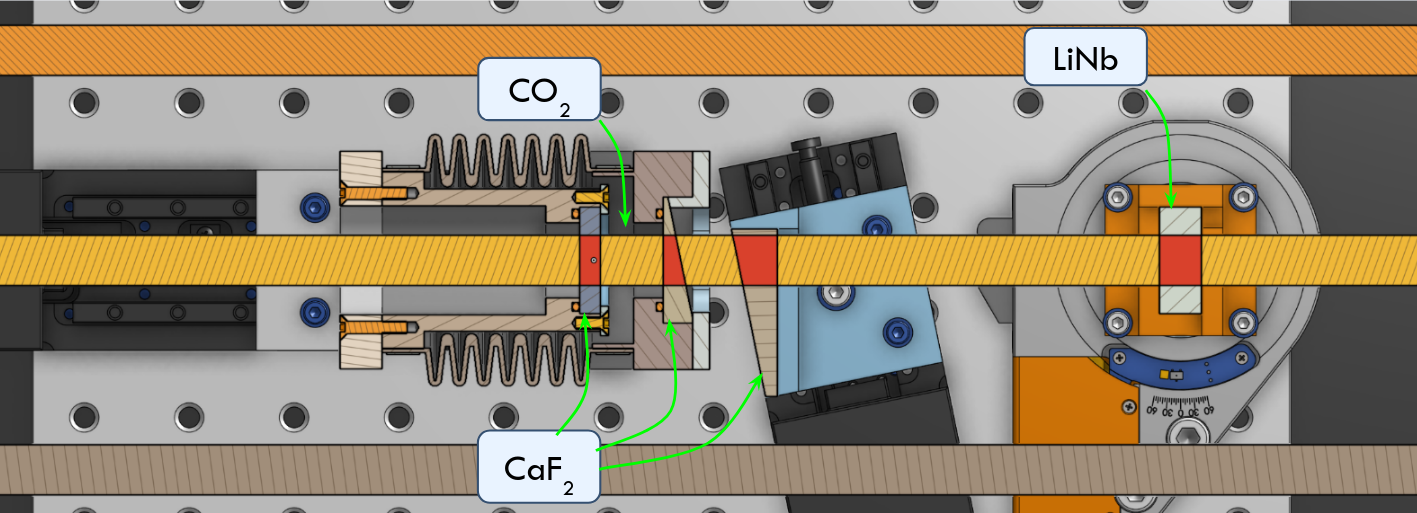}
        \caption{A CAD design of one LDC of NOTT (air delay line not represented). From left to right, the beam in question (yellow) passes through a simple wedged glass window entering into the variable length CO$_2$ chamber enclosed by bellows. The other end of the chamber is closed by the fixed prism of dispersing glass. A longer prism of the same material forms an equivalent variable thickness by translating along their angled face by 25mm. A flat LiNb glass plate on a rotating stage creates an adjustable birefringence term. The adjacent beams are shown in orange and gray, and their LDC will be offset longitudinally to manage space.}
        \label{fig:cad_ldc}
    \end{figure}

    One possible solution to correct for non-linear chromatic phase errors is an adaptive nuller such as investigated in the context of the TPF-I mission \cite{Lay2003, Peters2010}. This solution is however complex and expensive, exceeding the current budget of the NOTT project. An alternative solution identified is to include a variable length of CO$_2$ to balance the atmospheric contribution. A CO$_2$ concentration of 450 ppm in up to 100 m imbalanced path can be compensated with 45 mm of pure, ambient pressure CO$_2$. As shown in Fig. \ref{fig:cad_ldc}, the gas is kept between a flat, wedged window and the orthogonal surface of the static prism of the corrector. A silicone bellow system allows the frictionless travel of the wedged window over 50 mm on a translation stage.

    \begin{figure}
        \centering
        \includegraphics[width=0.9\textwidth]{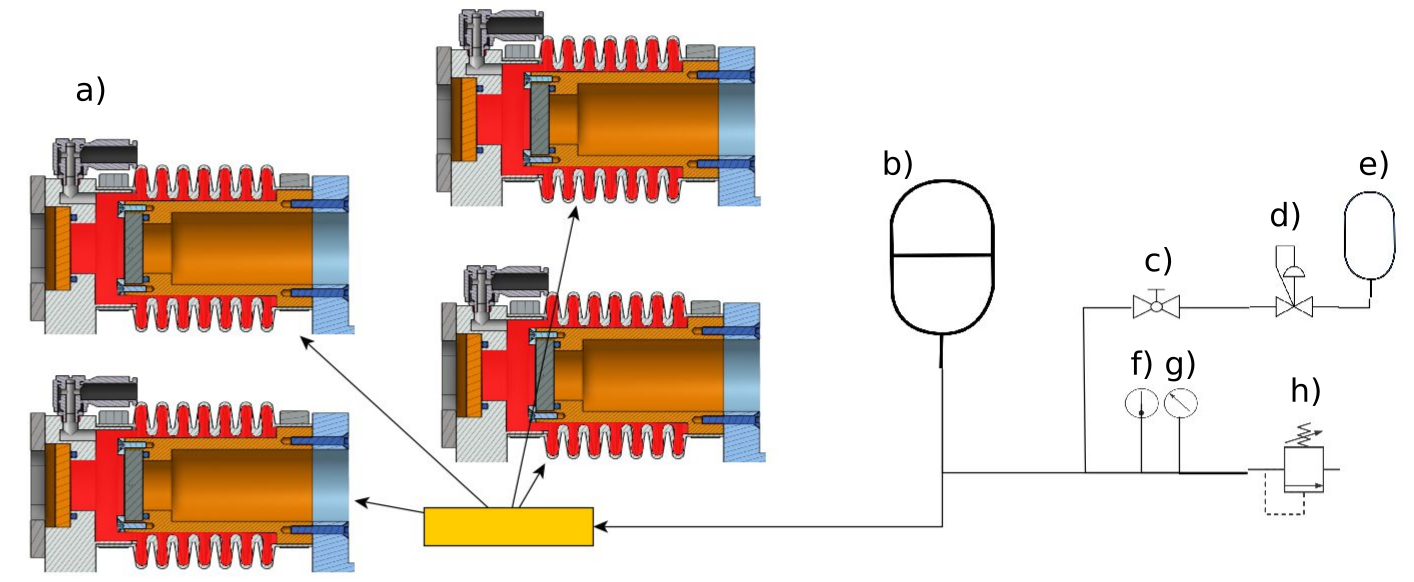}
        \caption{The pneumatic schematic of the gas system. The four chambers (a) are connected by tubes with each-other and with a 3L bladder (b) that ensures a constant, near-ambient pressure for all of them. While the goal is to coordinate the motion of the three chambers to maintain constant volume, but the bladder and a relief valve (h) are designed to prevent raise in pressure regardless. Filling the system is done with a portable pressure regulator (d) and small 16g cartridges (e), to avoid any risk during this maintenance operation. Temperature (f) and pressure (g) sensors allow for compensation of small variations in pressure by an adjustment in lengthe setpoint.}
        \label{fig:gaz_circuit}
    \end{figure}

    As shown in Fig. \ref{fig:gaz_circuit}, the four CO$_2$ chambers are connected by a network of tubes, allowing a coordinated motion maintaining a constant total volume. The system is designed to account for the volume changes brought by changes in ambient temperatures and motion problems (such as servo initialization), with minimal changes in pressure. This is done by connecting the system to a bladder protected in a box left at ambient pressure. A thermal sensor measures the temperature of the chambers and differential pressure sensor measures the pressure in the system. The residual pressure exerted by the bladder can be accounted for in the control algorithm.

    Filling the system with CO$_2$ is done using food-grade 16 g CO$_2$ cartridge mounted on an adjustable pressure regulator. The regulator is adjusted to a small pressure and the fill valve is partially open to limit the flow and fill speed. During this operation the influx of gas flushes the system towards the low-pressure relief valve that limits the pressure of the system, keeping the pressure sensor and any other sensitive components safe from pressure buildup throughout the whole lifecycle of the system.

\section{Glass selection}\label{sec:glasses}
    The NOTT LDC work primarily in the L band, but it must transmit at least a fraction of visible light that will be used for alignment with screens and the alignment monitoring cameras of NOTT. To be efficient, the glass must provide a large dispersion term $\partial n / \partial \lambda$ with a low mean refractive index in the infrared, so as to require minimal compensation in air path to obtain a given dispersion. It will also limit the pupil shift by the device in the infrared.
    
    An additional concern is the difference in beam shift obtained in the infrared compared to the visible at which the rough alignment of the beam on the cold stop will be done. To reduce this offset, the difference between visible and infrared indices must remain small.

    \begin{table}[]
        \centering
        \begin{tabular}[]{@{}llllll@{}}
            \hline
            Glass & n 0.7 µm & n 3.5 µm & n 4.0 µm & Pupil shift [\%] & $\phi^2$ [$rad^2$]\tabularnewline
            \hline
            ZnSe & 2.558 & 2.429 & 2.4167 & 8. & 1e-5 \tabularnewline
            CaF$_2$ & 1.43 & 1.4139 & 1.4095 & 0.5 &  1e-3\tabularnewline
            Al$_2$O$_3$ & 1.76314 & 1.694974 & 1.675619 & 0.5 & 1e-3 \tabularnewline
            Ge &  -  & 3.9901 & 3.9817 & - & - \tabularnewline
            ZnS & 2.33052 & 2.2554 & 2.25268 & - & - \tabularnewline
            \hline
        \end{tabular}

        \caption{Summary of some of the key properties of glasses considered for the dispersion correction. The pupil shift assumes a distance betweent the two prisms of 30 mm.}
        \label{tab:my_label}
    \end{table}

\begin{figure}
        \centering
        \includegraphics[width=0.8\textwidth]{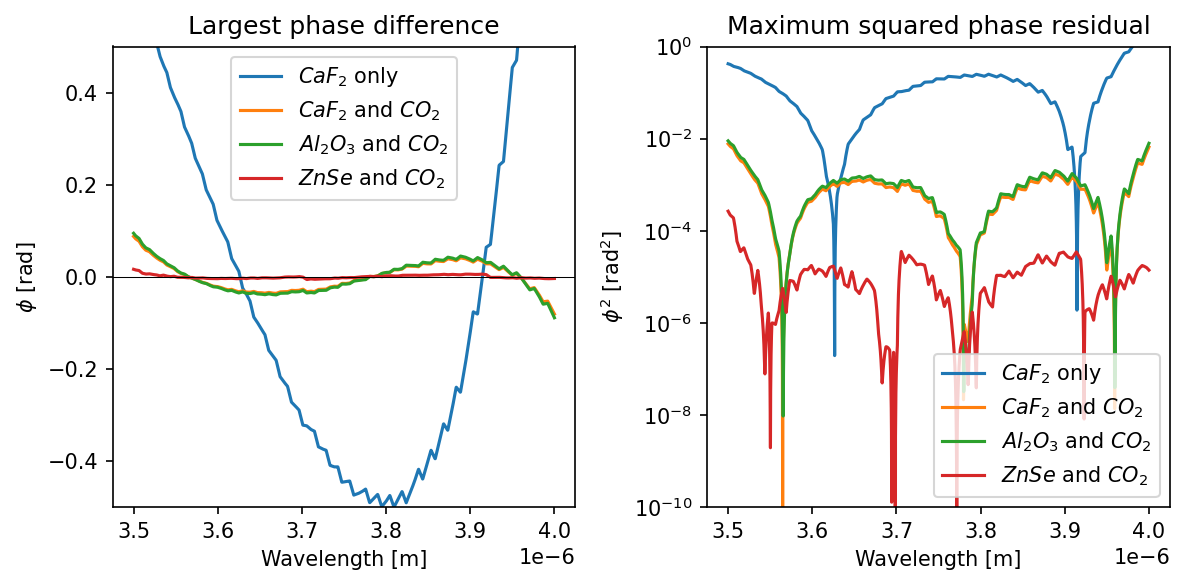}
        \caption{Residual phase (left) and squared phase (right) under different correction hypotheses and for the worst baseline. In glass-only situations, the CO$_2$ contribution leads to phase residuals of more than $500$ mrad. After correction with CO$_2$ the residual falls to $40$ mrad for CaF$_2$ and $1.6$ mrad for ZnSe.}
        \label{fig:residual}
    \end{figure}

    Figure \ref{fig:residual} shows the beam-to-beam residual after compensating with both a glass a and CO$_2$. CaF$_2$, Al$_2$O$_3$ and ZnSe all fill the requirement of a raw contrast of $10^{-2}$. While ZnSe offers a better performance in phase adjustments, several practical considerations make it less ideal, Its high refractive index (2.4 in L and 2.55 in R) in particular leads to the need for anti-reflection coating, and a larger beam shift in the visible, where the alignment will take place.

\section{Tuning strategy}
    The beam conditioning system of NOTT includes:
    \begin{enumerate}
        \item air delay lines,
        \item air-displacing glass,
        \item air-displacing CO$_2$,
        \item LiNb birefringence compensating plates.
    \end{enumerate}
    These create four degrees of freedom per beam. After excluding the common mode that is irrelevant in homodyne interferometry, this creates twelve independent degrees of freedom to adjust in order to create an optimal null. The result is measured on the four interferometric outputs measured independently in each polarization and for each spectral channels. This creates $ 8 n_{\lambda} $ measurements to inform on the result. However in the double-Bracewell configuration \cite{Angel1997}, the information on the input phase is degenerate by design.
    
    The strategy is to create a known piston offset probe in each of the beams to create a diversity in the outputs sensed. Measurements with just a few probes should enable the retrieval of the chromatic phase function, and compute the optimal position for each of the compensation actuators. This approach will be investigated and described in detail in coming publications.
    
\section{Conclusions and future work}
    The solution outlined here will allow NOTT to operate in the L band despite the pronounced phase feature of the asymmetric stretch of CO$_2$ centered at 4.26 µm, by adding a variable length of pure CO$_2$ at ambient conditions. Our design ensures safety of operation and maintenance. Different solutions in dispersing glass offer performance between $10^{-3}$ and $10^{-5}$ in raw contrast. Performance validation in the instrument will take place later in 2024 as part of preliminary integration in Leuven and Nice.
    
\acknowledgments 
This work has received funding from the Research Foundation -  Flanders (FWO) under the grant number 1234224N. SCIFY has received funding from the European Research Council (ERC) under the European Union's Horizon 2020 research and innovation program (grant agreement CoG - 866070).

\bibliography{nott_ldc} 
\bibliographystyle{spiebib} 

\end{document}